\documentclass[pmlr,twocolumn,10pt]{jmlr} % W&CP article

\usepackage{booktabs}
\usepackage{rotating}
\usepackage{siunitx}

\theorembodyfont{\upshape}
\theoremheaderfont{\scshape}
\theorempostheader{:}
\theoremsep{\newline}

\jmlrvolume{}
\jmlryear{}
\jmlrsubmitted{LEAVE UNSET}
\jmlrpublished{LEAVE UNSET}
\jmlrworkshop{Conference on Health, Inference, and Learning (CHIL) 2023} % W&CP title

\usepackage{xurl}

\usepackage{lipsum}  
\usepackage{mathtools}
\usepackage{amsmath,amsfonts,mathrsfs,amssymb, array, dsfont}
\usepackage{bm}
\usepackage{comment}

\newcommand{\bpi}{\bm \pi}

\newcommand{\bx}{\bm x}

\newcommand{\blambda}{\bm \lambda}
\newcommand{\balpha}{\bm \alpha}
\newcommand{\btheta}{\bm \theta}

\newcommand{\DD}{\mathcal{D}}

\newcommand{\SetS}{\mathcal{S}}
\DeclareMathOperator*{\argmax}{argmax}

\title[Bayesian Active Verbal Autopsy Questionnaire Design]{Bayesian Active Questionnaire Design for Cause-of-Death Assignment Using Verbal Autopsies}

\author{%
\Name{Toshiya Yoshida} \Email{toyoshid@ucsc.edu}\\
\addr University of California Santa Cruz, USA
\AND
\Name{Trinity Shuxian Fan} \Email{fansx@uw.edu}\\
\addr University of Washington, USA
\AND
\Name{Tyler McCormick} \Email{tylermc@uw.edu}\\
\addr University of Washington, USA
\AND
\Name{Zhenke Wu} \Email{zhenkewu@umich.edu}\\
\addr University of Michigan, USA
\AND
\Name{Zehang Richard Li} \Email{lizehang@ucsc.edu}\\
\addr University of California Santa Cruz, USA
}

\begin{document}

\maketitle

% This is the abstract for this article. If you are making your code
% available, \emph{do not link to it in the abstract since many indexing
% services will automatically remove or redact the link}. Instead,
% we are requiring every paper to have an initial statement on data and
% code availability right after the abstract.
\begin{abstract}
Only about one-third of the deaths worldwide are assigned a medically-certified cause, and understanding the causes of deaths occurring outside of medical facilities is logistically and financially challenging. Verbal autopsy (VA) is a routinely used tool to collect information on cause of death in such settings. VA is a survey-based method where a structured questionnaire is conducted to family members or caregivers of a recently deceased person, and the collected information is used to infer the cause of death. As VA becomes an increasingly routine tool for cause-of-death data collection, the lengthy questionnaire has become a major challenge to the implementation and scale-up of VA interviews as they are costly and time-consuming to conduct. In this paper, we propose a novel active questionnaire design approach that optimizes the order of the questions dynamically to achieve accurate cause-of-death assignment with the smallest number of questions. We propose a fully Bayesian strategy for adaptive question selection that is compatible with any existing probabilistic cause-of-death assignment methods. We also develop an early stopping criterion that fully accounts for the uncertainty in the model parameters. We also propose a penalized score to account for constraints and preferences of existing question structures. We evaluate the performance of our active designs using both synthetic and real data, demonstrating that the proposed strategy achieves accurate cause-of-death assignment using considerably fewer questions than the traditional static VA survey instruments.   
\end{abstract}

\paragraph*{Data and Code Availability}
% This initial paragraph is \textbf{mandatory}. Briefly state what data you
% use (including citations if appropriate) and whether the data are
% available to other researchers.\footnote{An example data availability
% statement: This paper uses the MIMIC-III dataset
% \citep{johnson2016mimic}, which is available on the PhysioNet repository
% \citep{johnson2016physionet}.}
% If you are not sharing code, you must explicitly state that you are not
% making your code available. If you are making your code available, then
% at the time of submission for review, please include your code as
% supplemental material or as a code repository link; in either case, your
% code must be anonymized. If your paper is accepted, then you should
% de-anonymize your code for the camera-ready version of the paper. \emph{If
% you do not include this data and code availability statement for your
% paper, or you provide code that is not anonymized at the time of
% submission, then your paper will be desk-rejected.} Your experiments later
% could refer to this initial data and code availability statement if it is
% helpful (e.g., to avoid restating what data you use).
The data and code to replicate this paper are publicly available. We use synthetic data generated with the replication codes and the Population Health Metrics Research Consortium (PHMRC) gold-standard VA dataset, which is publicly available at \url{https://ghdx.healthdata.org/record/ihme-data/population-health-metrics-research-consortium-gold-standard-verbal-autopsy-data-2005-2011}.

\paragraph*{Institutional Review Board (IRB)}
% This initial paragraph is \textbf{mandatory}. If your research requires IRB
% approval or has been designated by your IRB as Not Human Subject
% Research, then for the camera-ready version of the paper, you must
% provide IRB information (and at the time of submission for review, you
% can say that this IRB information will be provided if the paper is
% accepted). If your research does not require IRB approval, then you
% must state this to be the case. 
The study does not require IRB approval.

\section{Introduction}
\label{intro}
Data on cause of death is essential for understanding the heterogeneous burden of diseases. Most low- and middle-income countries, however, do not have vital statistics systems that produce high quality statistics on cause of death. As a result, about two-thirds of deaths worldwide are not registered or assigned a cause \citep{world2021civil}. Verbal autopsy (VA) is a widely used tool to collect information on cause of death when medically certified cause-of-death information is not available. VA is conducted through a structured questionnaire administered to family members or caregivers of a recently deceased person. The questionnaire collects information about the circumstances, signs, and symptoms leading up to a person's death. VAs are widely implemented in health and demographic surveillance systems, as well as national and multi-national mortality surveillance programs \citep[see, e.g., ][for an overview]{chandramohan2021estimating}. Data collected by VAs can be interpreted and assigned a cause of death by physician panels or, more commonly, analyzed by statistical algorithms. Algorithmic and statistical models have been developed and routinely used to classify individual cause of death and estimate the population-level cause-specific mortality fractions. The earlier and more widely used models typically assume symptoms are conditionally independent given causes and perform cause-of-death assignment using different variations of the Naive Bayes classifier \citep[e.g.,][]{byass2019integrated, mccormick2016}. Several extensions to \citet{mccormick2016}  have been recently proposed to further improve VA cause-of-death assignment using more complex Bayesian hierarchical models \citep[e.g., ][]{tsuyoshikunihama2020, li2020using, moran2021}. More flexible machine learning models have also been used in analyzing text-based narratives collected during VA interviews but it has been shown that they do not improve cause-of-death assignment using only the data from the structured questionnaire \citep{blanco2020extracting}. An overview of existing cause-of-death assignment algorithms can be found in \citet{li2022openva}.

Much of the literature analyzing VA data focuses on automating the process of classifying causes of death. The data collection process has received much less attention in the literature. One of the main barriers to scaling up VA implementation is the challenge of conducting overly complex and long questionnaires. For example, the current WHO 2016 standardized VA instrument includes $480$ questions. While each interview evokes only a subset of the questions, a typical VA interview still needs to go through $100$ to $200$ questions. Lengthy interviews increase the emotional stress to both the respondents and interviewers and can lead to survey fatigue and decreased acceptance of the interview \citep{loh2021added, nichols2022mixed, hinga2021ethical}. To the best of our knowledge, two attempts have been made in the last two decades to systematically reduce the length of VA questionnaires. \citet{serina2015shortened} measured the marginal associations between symptoms and causes of death and removed symptoms based on the ranking of their importance. However, they evaluated symptom importance based on a single highly simplified classification algorithm; thus the results are heavily influenced by the parametric assumptions of the algorithm. A more recent development to simplify the VA questionnaire was carried out by the WHO in producing the 2022 standard VA instrument, described in \citet{chandramohan2021estimating} and \citet{nichols2022mixed}. A more thorough process was carried out in which symptom response patterns and importance were evaluated in a model-agnostic fashion, and mixed-methods analyses were conducted to identify around $100$ questions that could be removed. 

Both previous approaches to shortening the VA interview are limited by the nature of the traditional survey instrument: the questionnaire needs to capture relevant symptoms associated with all potential causes of death.
% , and it is impossible to customize the questions and their order beyond the pre-defined flow of the questions. 
As the cause of death is the target of inference and unknown to the interviewer ahead of time, all questions need to be answered with the same priority by each respondent. Therefore, for a single interview, a large fraction of the collected data could potentially provide little relevant information in determining the cause of the particular death. 
% In addition to the inefficiency, the lengthy questionnaires likely lead to more data quality issues for downstream analysis, such as higher chance of non-response and measurement errors, as evidenced by survey research in different fields [CITE]. 

In this paper, we propose a statistical framework to adaptively conduct VA interviews by actively choosing the most informative questions to ask based on the collected responses and stopping when enough information has been collected to determine the cause of death. Unlike the static screening approach to reduce the length of the questionnaire for all respondents, our approach leads to a dynamic and individualized questionnaire design that is optimized for the classification of each death. Our approach is motivated by the methods developed in the field of active learning and computerized adaptive testing. We focus our attention to its application in verbal autopsy questionnaires, as this is the first time an adaptive design is considered for the purpose of verbal autopsy questionnaires. The proposed approach, however, can also be generally useful for surveys with the aim of classifying respondents into pre-defined groups. The main contributions of our work can be summarized as follows:
\begin{enumerate}
    \item Our active question selection strategy optimizes the cause-of-death classification of each individual death dynamically. We demonstrate that for most deaths, a small number of actively selected questions is enough to achieve the same level of classification accuracy as when using all questions. 
    \item Our active questionnaire design is compatible with any existing probabilistic cause-of-death assignment algorithms and thus can be applied regardless of the choice of analysis models used to describe the joint distribution of symptoms and causes. This allows more flexibility in practice, as an analyst can choose the most appropriate analysis model to conduct the cause-of-death assignment and seamlessly adopt the proposed active questionnaire design for data collection.
    \item Our approach extends existing work in psychometrics to a more principled Bayesian strategy that fully accounts for the uncertainty of the cause-of-death classification model, and we show that it leads to uncertainty-aware stopping rules that are more appropriate for high-stake tasks such as VAs.
    \item We also propose a novel penalized version of the adaptive questionnaire strategy to account for practical constraints and preferences for the order of the questions.
\end{enumerate}

\section{Preliminaries}
\subsection{Active Learning}
Active learning has been studied extensively in many areas of machine learning. Active learning algorithms seek to choose the optimal data instances to be used for the learning system. Most of the work in active learning focuses on choosing data points to be labeled to improve the performance of classification algorithms \citep{settles2011theories}. Many types of data query strategies have been proposed in the literature, including uncertainty sampling \citep{lewis1995sequential}, query-by-committee \citep{seung1992query}, and approaches that aim to reduce the variance \citep{cohn1996active} and generalization errors \citep{roy2001toward}. More similar to the context of this paper, active learning approaches have also been used to query complete feature vectors within a pool of observations with missing values \citep{melville2004active, li2021active}. Active learning has been successfully applied to natural language modeling \citep{kaushal2019learning}, computer vision \citep{dor2020active}, and many other applications. Previous work on active learning to collect survey responses is scarce. The work most related to our approach is the active matrix factorization approach for surveys measuring voter opinion proposed in \citet{zhang2020active}. They developed an active question selection strategy to optimize the estimation of latent profiles of respondents under a low-rank matrix factorization model.

\subsection{Computerized Adaptive Testing}
% CD-CAT, SHE, PWKL
% The SHE algorithm which was proposed by \cite{tatsuoka2002}.
% This strategy chooses the next item so that the expected Shannon entropy is minimized.

Computerized adaptive testing (CAT) is a mode of testing that aims to find the optimal set of questions for each individual, thus resulting in more efficient and accurate recovery of latent traits of examinees ~\citep{weiss1984application}. CAT was originally proposed for item response theory (IRT) in~\cite{lord1971robbins}. The items are selected to maximize the test information at the current estimated ability based on IRT from an item bank. One main application of CAT is in the cognitive diagnosis models (CDMs), which is termed as cognitive diagnostic computerized adaptive testing (CD-CAT) ~\citep[see, e.g., ][]{cheng2009,huebner2010overview}. As CDMs have a discrete attribute space, standard CAT approaches for IRT are not directly applicable to CDMs. Several CD-CAT methods have been proposed in the literature with different item selection criteria. Two of the most widely adopted class of methods are the Shannon entropy approach, proposed by \cite{tatsuoka2002} and \citet{tatsuoka2003}, and various procedures based on the Kullback-Leibler (KL) information \citep[e.g.][]{xu2003simulation, cheng2009}. Other information metrics including mutual information~\citep{wang2013mutual} and large deviation~\citep{liu2015rate} are also proposed for CD-CAT. 
We will utilize a similar strategy based on KL information for selecting optimal questions for VA surveys in this paper.

\section{Method}

\subsection{Bayesian Active Questionnaire Design}
We assume that there exists a question bank with $J$ questions. 
% the goal of the active questionnaire design is to dynamically select the next question that produce the most information to infer the unknown cause of death. 
Let $X_{ij}$ denote the response to question $j$ for death $i$, $X_i=\{X_{i1},\dots,X_{ip}\}$ denote the vector of responses for death $i$, and $Y_i \in \{1, ..., C\}$ denote the categorical variable indicating the cause of death. We consider $X_{ij} \in \{0, 1\}$ in this paper since most of the questions collected by VA surveys are binary. The extension to general $X$ is straightforward and does not change the active design formulation. We consider the situation where a probabilistic model was fitted on a dataset $(X_i, Y_i)_{i = 1, ..., n}$ and produced estimates for the distribution $p(X, Y)$.

% In the context of CD-CAT, there are a few options to select the next question iteratively.
% The first one is 
% When we consider the survey responses having 0-1 responses, we choose the next item $j\notin S^{(t)}$ at iteration $t\in 1,\dots,p$ so that the following score is minimized
% \begin{align*}
%     Score_j &= \sum_{q=0}^1 \left( H(\{X_{ij}:j\in S^{(t)}\}, x_{ij}=q) \right.\\
%             & \quad \left. p(x_{ij}=q \mid \{X_{ij}:j\in S^{(t)}) \right)
% \end{align*}
% where 
% \begin{align*}
%     H(\{X_{ij}:j\in S^*\})  &= -\sum_{i=1}^C \left( p(Y_i=c \mid \{X_{ij}:j\in S^*\}) \right. \\
%                             & \quad \left. \log(p(Y_i=c\mid \{X_{ij}:j\in S^*\})) \right)
% \end{align*}
% and $S^*$ is the set of questions collected.

Our approach follows the KL information procedures in the CD-CAT literature \citep{cheng2009}. In the context of VA, after $t$ questions have been asked, we let $\SetS_t$ denote the set of questions already asked and among the questions $j \not\in \SetS_t$, we identify the question with the most different distribution under the current estimated cause of death compared to alternative causes. That is, for an alternative cause $y$ and the $j$-th question, we define
\begin{align*}
    D_j(\hat y_i^{(t)} \parallel y) 
    &= \sum_{x} q_{j}( x \mid \hat y_i^{(t)})\log\left( \frac{q_{j}(x \mid \hat y_i^{(t)})}{q_{j}(x\mid y)}\right),
\end{align*}
where $q_{j}(x \mid y) =  p(X_{ij}=x\mid Y_i = y)$ is the conditional distribution of the $j$-th indicator given the cause of death being $y$, and $\hat y_i^{(t)} = \argmax_y p(Y_i=y\mid \{X_{ij}:j\in \SetS_t\})$ is the estimated cause of death given the collected information at step $t$. Several different methods have been proposed to combine the KL distances to all alternative classifications in the CD-CAT literature. We adopt the idea of posterior weighted KL (PWKL) algorithm \citep{cheng2009} and maximize the weighted score for each question $j$ defined by
\begin{align}
    &\mbox{Score}_j = \sum_{y=1}^C  D_j(\hat y_i^{(t)} \parallel y)p(Y_i=y\mid \{X_{ij}:j\in \SetS_t\}).  \label{eq;PWKLScore}
\end{align}
The existing CD-CAT literature typically assumes that the conditional distributions involved in computing the scores are known or can be estimated with high precision from existing data \citep{chang2019nonparametric}. In the context of VA, however, these quantities need to be estimated using a cause-of-death assignment model with usually limited training data $\DD$. Several Bayesian methods have been introduced to infer cause of death using VAs \citep{mccormick2016,tsuyoshikunihama2020,li2021bayesian,wu2021double} and it has been shown that considerable uncertainties exist in the classification and parameter estimations in these models. To account for the full posterior uncertainty of the conditional probabilities used to construct the PWKL score, we instead propose the following posterior predictive PWKL score
\begin{align*}
    \mbox{PScore}_j    &= \int \mbox{Score}_j(\phi) p(\phi \mid \DD) d\phi \\
                &\approx \frac{1}{B} \sum_{b=1}^B \mbox{Score}_j(\phi^{(b)}), 
                % &= \sum_{b=1}^B\sum_{c=1}^C \left( D_j(Y_i^{(t)} \parallel Y_i=c, \phi^{(b)}) \right.\\
                % &\quad \left. p(Y_i=c \mid \{X_{ij}:j\in S^{(t)}\},\phi^{(b)}) \right)
\end{align*}
where we use $\phi$ to denote all parameters used in the assignment model, $\phi^{(b)}$ to denote the $b$-th draw of $\phi$ from the posterior distribution $p(\phi | \DD)$, $B$ to denote the number of draws, and $\mbox{Score}_j(\phi^{(b)})$ is the PWKL score defined in \equationref{eq;PWKLScore} with $\phi^{(b)}$ plugged in. In the rest of the paper, we refer to the active question selection strategy that maximizes $\mbox{Score}_j(\hat\phi)$,  where $\hat\phi$ is the posterior mean of $\phi$, as the design using point estimates and the strategy maximizing $\mbox{PScore}_j$ as  the design using posterior predictive scores.
We note that this is a further extension that accounts for the full model uncertainty, compared to the modified PWKL score proposed in \citet{kaplan2015}, where only the uncertainty of the latent classification was integrated over.

% The previous PWKL does not take into account the full posterior distribution of $Y_i^{(y)}$.
% This motivated the modified PWKL (MPWKL) score \cite{kaplan2015}.
% However, these methods treat the conditional distributions as known and thus not accounts for the uncertainty in any of the conditional probabilities.
% We can instead construct posterior predictive scores to replace \eqref{eq;PWKLScore}
% \begin{align}
%     PScore_j    &= \int Score_j(\phi) p(\phi \mid data)\\
%                 &= \frac{1}{B} \sum_{b=1}^B Score_j(\phi^{(b)})\\
%                 &= \sum_{b=1}^B\sum_{c=1}^C \left( D_j(Y_i^{(t)} \parallel Y_i=c, \phi^{(b)}) \right.\\
%                 &\quad \left. p(Y_i=c \mid \{X_{ij}:j\in S^{(t)}\},\phi^{(b)}) \right)
% \end{align}
% where $\phi$ is the set of parameters $\{\btheta, \bpi\}$, $\phi^{(b)}$ samples from the posterior distribution of $\phi$, and $B$ size of sampels drawn from the posterior distribution.

\subsection{Cause-of-Death Assignment Model}
\label{sec:model}
The proposed active selection strategy does not depend on any particular choice of the cause-of-death assignment model used to analyze the data, as long as the conditional probabilities in \equationref{eq;PWKLScore} can be computed. In this paper, we consider a simplified version of the algorithm proposed in \citet{mccormick2016} to analyze training dataset $\DD$. Our analysis model assumes the following data generating process
\begin{align*}
    Y_i                         &\sim {\rm Cat}(\bpi),\notag\\
    p(X_i=x_i \mid Y_i=y)   &= \prod_j \theta_{yj}^{x_{ij}}(1-\theta_{yj})^{1-x_{ij}},\label{eq;DGP}
\end{align*}
% where $f(\bx_i, c; \btheta)$ is the likelihood function.
% The simple example is $f(\bx_i, c; \btheta)=\prod_j \theta_{cj}(1-\theta_{cj})^{1-x_{ij}}$.
with conjugate priors $\theta_{yj}\sim\mbox{Be}(a_y, b_y)$ and $\bpi\sim\mbox{Dir}(\balpha)$ respectively. The posterior distributions of the parameters are
\[ \bpi\mid \DD \sim \mbox{Dir}(n_1 + \alpha_1,\dots,n_C + \alpha_C), \]
\[ \theta_{yj}\mid \DD \sim \mbox{Be}\left(a_y+\sum_{i:Y_i=y}x_{ij}, ~ b_y+n_y-\sum_{i:Y_i=y}x_{ij}\right),  \]
where $n_y=\sum_{i=1}^n \bm 1(Y_i=y)$ for $y=1,\dots, C$. When some of the $Y_i$ are unknown in the data, this assignment model can also be trivially extended to generate posterior draws of $(\bpi, \btheta, Y_{miss})$.

\subsection{Stopping Criterion}
In adaptive testing settings where the test duration is the main constraint, it is usually reasonable to stop the test after a pre-specified number of questions \citep[e.g.][]{chen2012, wang2011, cheng2009}. A fixed length stopping rule is straightforward to implement in our active VA questionnaire design as well. However, it is usually more appropriate to stop the interview only when enough precision has been achieved. 
Several related approaches to early stopping were developed in the literature. 
% \citet{choi2010} proposed a predicted standard error reduction (PSER) stopping rule based on the reduction in standard error \red{of what?} resulting from the additional question.
\citet{tatsuoka2002} proposed a stopping criteria where the largest probability of the classified class reached a given value, which was later adapted by \citet{hsu2013} where another condition on the second largest probability was added. In our notation, the criterion proposed in \citet{hsu2013} suggests stopping the questionnaire when the largest value of $p(Y_i=y\mid \{X_{ij}:j\in \SetS_t\})$ is larger than  $p_{1{\rm st}}$ and the second largest value is smaller than $p_{2{\rm nd}}$ where $p_{1{\rm st}}\ge p_{2{\rm nd}}$ are two pre-specified thresholds.

% The stopping rule of Hsu et~al \citep{hsu2013} is based on posterior probability of cause-of-death assignment (PPCDA) and outlined as follows:
% \begin{enumerate}
%     \item Prespecify $p_{1{\rm st}}$ and $p_{2{\rm nd}}$ where $p_{1{\rm st}}\ge p_{2{\rm nd}}$ (e.g., $p_{1{\rm st}}=0.9$ and $p_{2{\rm nd}}=0.1$). 
%     \item Stop the survey when the largest PPCDA is not smaller than $p_{1{\rm st}}$ and the second largest PPCDA is not greater than $p_{2{\rm nd}}$.
% \end{enumerate}

The simplicity of this stopping rule is appealing, but when parameters are estimated with large uncertainty, stopping the questionnaire based on point estimates of classification probabilities may lead to erroneous early stopping. Instead, we compute the posterior predictive probability of meeting a pre-specified stopping rule similar to that introduced in \citet{hsu2013}. At each iteration, for the $b$-th posterior draw $\phi^{(b)}$, we compute $p_y^{(b)} = p(Y_i=y\mid \{X_{ij}:j\in \SetS_t\}, \phi^{(b)})$. The current most likely cause assignment $y^\star$ is then the mode of $\{\argmax_{y=1, ..., C}p_y^{(b)}\}_{b=1,\dots, B}$. We can compute the posterior predictive probability for the event 
\begin{align*}
p(Y_i=y^\star\mid \{X_{ij}:j\in \SetS_t\}) &> p_{1{\rm st}},   \\ 
p(Y_i=y\mid \{X_{ij}:j\in \SetS_t\}) &< p_{2{\rm nd}},\;\;\;\forall y \neq y^{\star}.
\end{align*} 
We can then stop the survey when this probability exceeds a certain tolerance threshold $r \in (0, 1)$. More generally, while we adopt this specific stopping criterion in this paper, the fully Bayesian nature of the proposed score formulation also allows other stopping criteria to be similarly plugged in. 
% Denote the second most likely assignment associated with $b$-th posterior draw to be $\tilde y^{(b)} = \argmax_{y \neq y^\star}p_y^{(b)}$
% stop the survey when $100r\% $ of $\hat{p}^{(s)}_{c^*}$ are not smaller than $p_{1{\rm st}}$ and $100r\% $ of $\hat{p}^{(s)}_{c_2^{(s)}}$ are not grater than $p_{2{\rm nd}}$ where $\alpha\in[0,1]$ is a prespecified value.

% It would make sense to use each posterior draw of $\btheta$ and $\bpi$. 
% By using each $\btheta^{(s)}$, we can gain the order and corresponding PPCDA $\hat{p}_c^{(s)}$ for $s=1,\dots, N$ and $c=1,\dots,C$.
% In each draw, there is a cause $c$ that maximizes PPCDA $\hat{p}_c^{(s)}$ so we can define $c^{*(s)}=\argmax_c \hat{p}_c^{(s)}$.
% We can choose the mode of $c^{*(s)}$ as a classified class and denote it as $c^*$ and denote the PPCDA for this class in each draw as $\hat{p}^{(s)}_{c^*}$. 
% In each draw, we denote the cause with the largest PPCDA in the causes except for $c^*$ as $c_2^{(s)}$ and denote the corresponding PPCDA as $\hat{p}^{(s)}_{c_2^{(s)}}$.
% Note that, the $c^*$ is fixed while $c_2^{(s)}$ varies among the draws.
% We can stop the survey when $100r\% $ of $\hat{p}^{(s)}_{c^*}$ are not smaller than $p_{1{\rm st}}$ and $100r\% $ of $\hat{p}^{(s)}_{c_2^{(s)}}$ are not grater than $p_{2{\rm nd}}$ where $\alpha\in[0,1]$ is a prespecified value.

\subsection{Accounting for Existing Flow of the Questionnaire}
\label{sec:jump}
For traditional static VA surveys, the questionnaire structure usually follows a carefully designed order that leads to a natural flow of questions. The dynamic nature of the active questionnaire design inevitably breaks such an ordering of questions and may lead to consecutive questions that are concerned with very different aspects of an individual's death. While this is desired from the perspective of maximizing the collected information quickly, one practical concern for jumping across different topics is that it may increase the chance of inaccurate responses from the respondents. 
It is straightforward to impose deterministic skip patterns by modifying the search space $\SetS_t$ at each iteration based on the collected responses. For example, VA surveys typically include questions that are only triggered when a root-question has been answered. We may let $\SetS_0$ include only root-questions, and any sub-questions are added to $\SetS_t$ only after the corresponding root-question has been answered. 
In addition, it may also be useful to maintain some of the natural ordering of the questions in a `soft' fashion. The proposed active design strategy can be easily extended to incorporate such preference by adding a penalization for certain moves across the questions, i.e., let $j^{(t)}$ denote the index of question asked in the $t$-th iteration and define the penalized score 
\[ \mbox{PScore}_j' = \mbox{PScore}_j -  \lambda D(j, j^{(t)}), \]
where $\lambda>0$ is a parameter that regulates the degree of penalization for the jumping behavior and $D(j, k)$ is a pre-specified distance metric between the $j$-th and $k$-th questions. For example, when there is a group of questions that we would like to ask together but have no preference for the order within the group, we may let $D(j, k) = 0$ if the questions are within the same group and $1$ otherwise. 

\section{Experiments}
% To assess the feasibility and performance of the proposed active questionnaire design, we conduct several realistic simulation studies and provide an extensive case study using the Population Health Metrics Research Consortium (PHMRC) gold-standard VA dataset.

\subsection{Synthetic Data}
We first generate synthetic data to evaluate the performance of the active questionnaire design strategy. We consider the following two data generating processes:
\begin{enumerate}
    \item Correctly specified model: we generate observations using the model described in Section \ref{sec:model}. We let $C = 10$, $J = 50$, $\balpha = (1, ..., 1)$ and $(a_y, b_y)$ to be $(0.5,0.5)$ for $y = 1,2,3$, $(3,3)$ for $y = 4,5,6$, and $(1,3)$ for $y = 7,8,9,10$.
    \item Misspecified model: we generate observations by a latent class model such that each cause of death consist of multiple unobserved sub-categories. That is, we generate $Z_i \mid Y_i = y \sim {\rm Cat}(\blambda_y)$ and 
    \[
    p(X_i=\bx_i | Y_i=y, Z_i = k) = \prod_j \theta_{ykj}^{x_{ij}}(1-\theta_{ykj})^{1-x_{ij}}.
    \]
    We let $\blambda_c = (\lambda_{y1}, \lambda_{y2}, \lambda_{y3}) \sim \mbox{Dir}(1, 1, 1)$ and $\theta_{ykj} \sim \mbox{Beta}(1, 1)$ for all $y, k$, and $j$.
\end{enumerate}
In each case, we generate $n_1 = 200$ and $1000$ training observations, respectively, and evaluate the performance of different questionnaire designs on $n_0 = 200$ test observations. 

\begin{figure}[htbp]
\floatconts
  {fig:nonstop}
  {\caption{Classification accuracy of different questionnaire design as the number of questions increases.}}
  {\includegraphics[width=\linewidth]{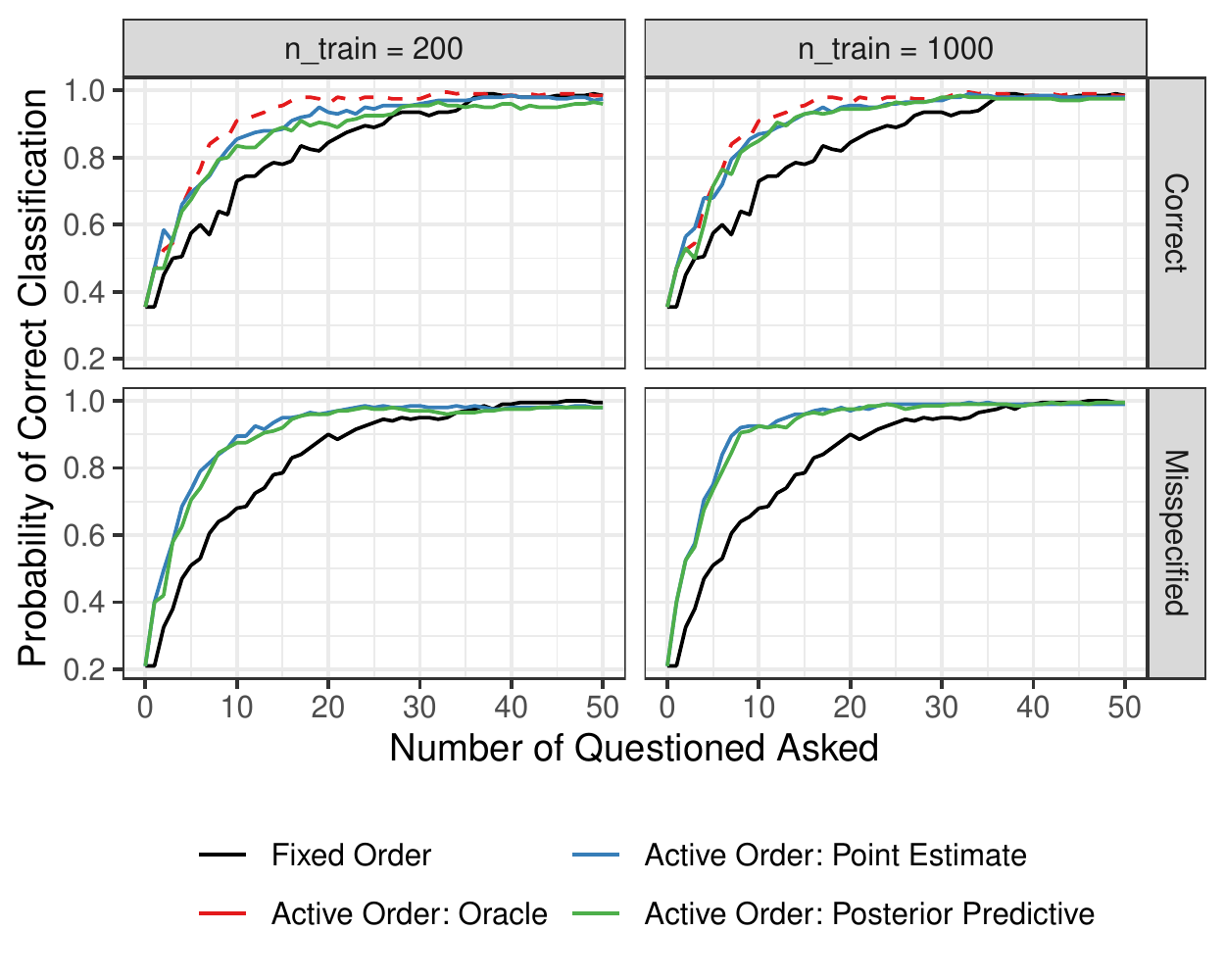}}
\end{figure}

In the first experiment, we consider running the questionnaire with a fixed number of questions. Figure \ref{fig:nonstop} shows the probability of correct classification of cause of death in the test set, given different lengths of the questionnaire. In all cases, the two active question selection strategies reach high classification accuracy faster than asking the questions sequentially with a fixed order. In the case of the correctly specified model, we also evaluate the accuracy of the oracle strategy when the parameters of the true data generating process are known, and both active strategies perform similarly to the oracle when the training data is sufficiently large.

\begin{table}[htb]
\setlength{\tabcolsep}{3pt}
\floatconts
  {tab:stopping1}
  {\caption{Classification accuracy and questionnaire length for the synthetic data under the correctly specified model. Median and $5$th and $95$th percentiles of the questionnaire length are shown for each stopping rule condition.}}
  { \vspace{-.3cm}
  \begin{tabular}{cclcccc}
  \toprule
   $p_{1st}$ &  $d$ & Stopping Rule & Acc & Median &Lower & Upper\\\midrule 
   $0.8$ & $0.5$ & Point Est & $0.85$ & $5$ & $3$ & $14$\\
          && Pred $r = 0.5$ & $0.92$& $10$& $4$& $50$\\
          && Pred $r = 0.7$ & $0.95$& $14$& $6$&$50$\\
   \midrule
   $0.8$ & $0$ & Point Est & $0.96$ & $8$ & $5$ & $35$\\
          && Pred $r = 0.5$ & $0.96$ & $13$& $6$& $50$\\
          && Pred $r = 0.7$ & $0.96$& $20$& $7$& $50$\\
   \midrule
   $0.9$ & $0.5$ & Point Est & $0.95$ & $7$ & $5$ & $18$\\
          && Pred $r = 0.5$ & $0.95$& $12$& $6$&$50$\\
          && Pred $r = 0.7$ & $0.96$& $18$& $7$&$50$\\
   \midrule
   $0.9$ & $0$ & Point Est & $0.97$ & $10$ & $7$ & $50$\\
          && Pred $r = 0.5$ & $0.96$& $16$& $8$&$50$\\
          && Pred $r = 0.7$ & $0.96$& $23$& $8$&$50$\\
    \bottomrule
  \end{tabular}}
\end{table}

\begin{table}[htb]
\setlength{\tabcolsep}{3pt}
\floatconts
  {tab:stopping2}
  {\caption{Classification accuracy and questionnaire length for the synthetic data under the misspecified model. Median and $5$th and $95$th percentiles of the questionnaire length are shown for each stopping rule condition.}}
  {  \vspace{-.3cm}
    \begin{tabular}{cclcccc}
  \toprule
   $p_{1st}$ &  $d$ & Stopping Rule & Acc & Median &Lower & Upper\\\midrule 
   $0.8$ & $0.5$ & Point Est & $0.88$ & $5$ & $3$ & $11$\\
          && Pred $r = 0.5$ & $0.88$& $4$& $3$& $18$\\
          && Pred $r = 0.7$ & $0.94$& $5$& $3$&$27$\\
   \midrule
   $0.8$ & $0$ & Point Est & $0.96$ & $6$ & $5$ & $17$\\
          && Pred $r = 0.5$ & $0.94$ & $6$& $4$& $30$\\
          && Pred $r = 0.7$ & $0.99$& $8$& $5$& $50$\\
   \midrule
   $0.9$ & $0.5$ & Point Est & $0.94$ & $6$ & $4$ & $13$\\
          && Pred $r = 0.5$ & $0.96$& $6$& $4$&$23$\\
          && Pred $r = 0.7$ & $0.98$& $7$& $4$&$50$\\
   \midrule
   $0.9$ & $0$ & Point Est & $0.96$ & $7$ & $6$ & $20$\\
          && Pred $r = 0.5$ & $0.98$& $8$& $5$&$42$\\
          && Pred $r = 0.7$ & $0.99$& $9$& $5$&$50$\\
    \bottomrule
  \end{tabular}}
\end{table}

We then evaluate the performance of the active design under different varying-length stopping rules under the same data generating processes as in the first experiment. We compare the classification accuracy of different active design strategies under the two data generating processes, respectively, when different stopping criteria are satisfied. We consider three stopping rules: when using the point estimates to select questions, we consider the same criterion in \citet{hsu2013}; when using the posterior predictive scores, we consider stopping when $r = 50\%$ and $70\%$ of the posterior draws meet the same criterion. Tables \ref{tab:stopping1} and \ref{tab:stopping2} summarize the results. We fix $p_{1{\rm st}} = 0.8$ and $0.9$, and for the second threshold, we compute $p_{2{\rm nd}} = (1 - p_{1{\rm st}}) / C + d(C - 2)(1 - p_{1{\rm st}})/C$ for various choices of $d \in [0, 1]$, as suggested by \citet{hsu2013}. Overall, classification accuracy increases as we increase $p_{1{\rm st}}$ and decrease $ p_{2{\rm nd}}$ (or equivalently decrease $d$), as more questions need to be asked before the more stringent stopping criterion are met.
Due to the space limitation, we present results for $d = 0$ and $0.5$ only. In addition to accuracy, we also compare the median and $5$th and $95$th percentiles of the questionnaire length. It is worth noting that in Table \ref{tab:stopping2}, when the analysis model is misspecified, the stopping rule with $70\%$ posterior probability of satisfying the stopping criterion leads to similar median questionnaire length compared to considering only point estimates, while also having a longer tail in the number of questions asked. This leads to higher overall accuracy. In high-stakes contexts such as VA, it is often more appropriate to favor conservative strategies and collect more information for deaths that are difficult to classify, especially when the analysis model driving the questionnaire design is not accurate enough. In such cases, the proposed stopping rule with a larger $r$ might be preferred in practice.

Finally, we consider situations where the responses may be subject to errors when the questionnaire deviates from a natural flow. Let $j^{(t)}$ denote the index for the $t$-th question asked, and we simulate noisy responses during the interview by letting  
\[ 
X_{ij^{(t)}}^\star = \begin{cases}  1 - X_{ij^{(t)}} & \mbox{ w.p. $\frac{D(j^{(t-1)}, j^{(t)})}{h}$}, \\ X_{ij^{(t)}} & \mbox{ w.p. $1 - \frac{D(j^{(t-1)}, j^{(t)})}{h}$}. \end{cases} 
\]

Here we consider the distance metric $D(j, k) = |j - k|/p$. In this setup, asking questions sequentially leads to the least amount of added noise. We compare different active questionnaire designs using the penalized score described in Section \ref{sec:jump} with $\lambda = 2$ and $10$. We consider the same data generating process as in the first experiment, and both a low noise setting ($h = 10$) and a high noise setting ($h = 2$). Figure \ref{fig:jump} shows that the unpenalized active designs lead to sub-optimal classification accuracy in high noise settings, as the responses include more errors induced by the non-sequential order of the questions. Active designs based on the proposed penalized score are able to mitigate the effect of response errors and achieve classification accuracy comparable to or higher than static questionnaire designs with the optimal order. 

 \begin{figure}[htb]
\floatconts
  {fig:jump}
  {\caption{Classification accuracy of different questionnaire design when responses include order-induced noises. The red curves correspond to fixed sequential order.}}
  {\includegraphics[width=\linewidth]{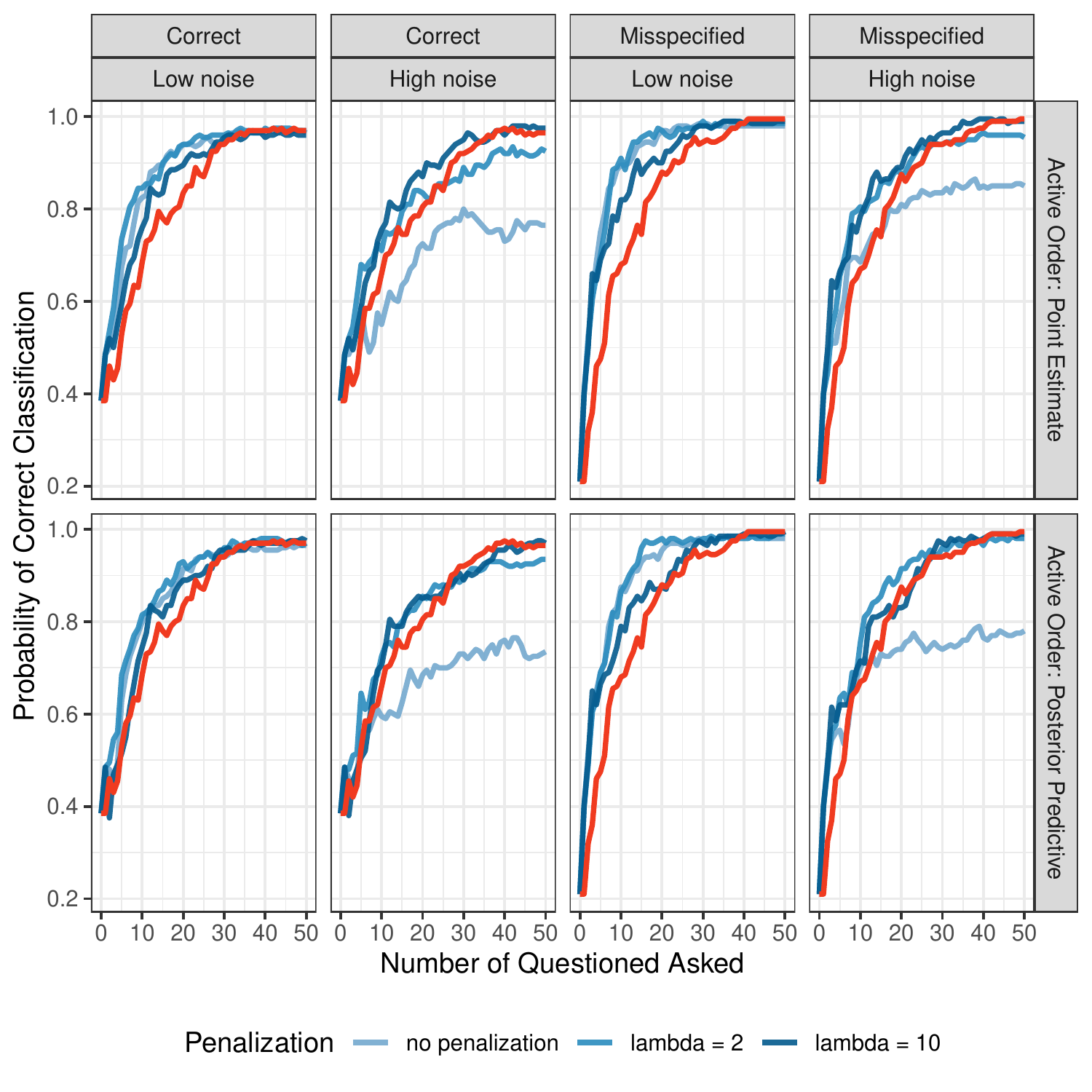}}
\end{figure}

\subsection{PHMRC Data}

In this section, we consider the application of the adaptive questionnaire strategy on the Population Health Metrics Research Consortium (PHMRC) gold-standard VA dataset \citep{murray2011population}. The PHMRC dataset is widely used for validating VA cause-of-death assignment methods \citep{mccormick2016, tsuyoshikunihama2020, moran2021}. It consists of $7,841$ adult deaths collected from six study sites (Andhra Pradesh, India; Bohol, Philippines; Dar es Salaam, Tanzania; Mexico City, Mexico; Pemba Island, Tanzania; and Uttar Pradesh, India). Gold-standard causes were determined based on laboratory, pathology, and medical imaging findings. In this dataset, there are $C = 34$ cause-of-death categories, and we pre-processed the raw dataset into $J = 168$ binary indicators using the steps described in \citet{li2022openva}.

We again consider both stopping at a fixed length and when using the varying-length stopping rules. We conduct a 10-fold cross-validation analysis, where we evaluate different active questionnaire strategies on each fold of data with the rest of the data being used to estimate model parameters. We treat missing values in the dataset to be missing at random when fitting the model, as is commonly assumed in existing VA cause-of-death assignment algorithms \citep{mccormick2016, tsuyoshikunihama2020, li2021bayesian}.

\begin{figure}[hbt]
\floatconts
  {fig:phmrc1}
  {\caption{Classification accuracy of different questionnaire designs on the PHMRC data as the number of questions increases.}}
  {\includegraphics[width=\linewidth]{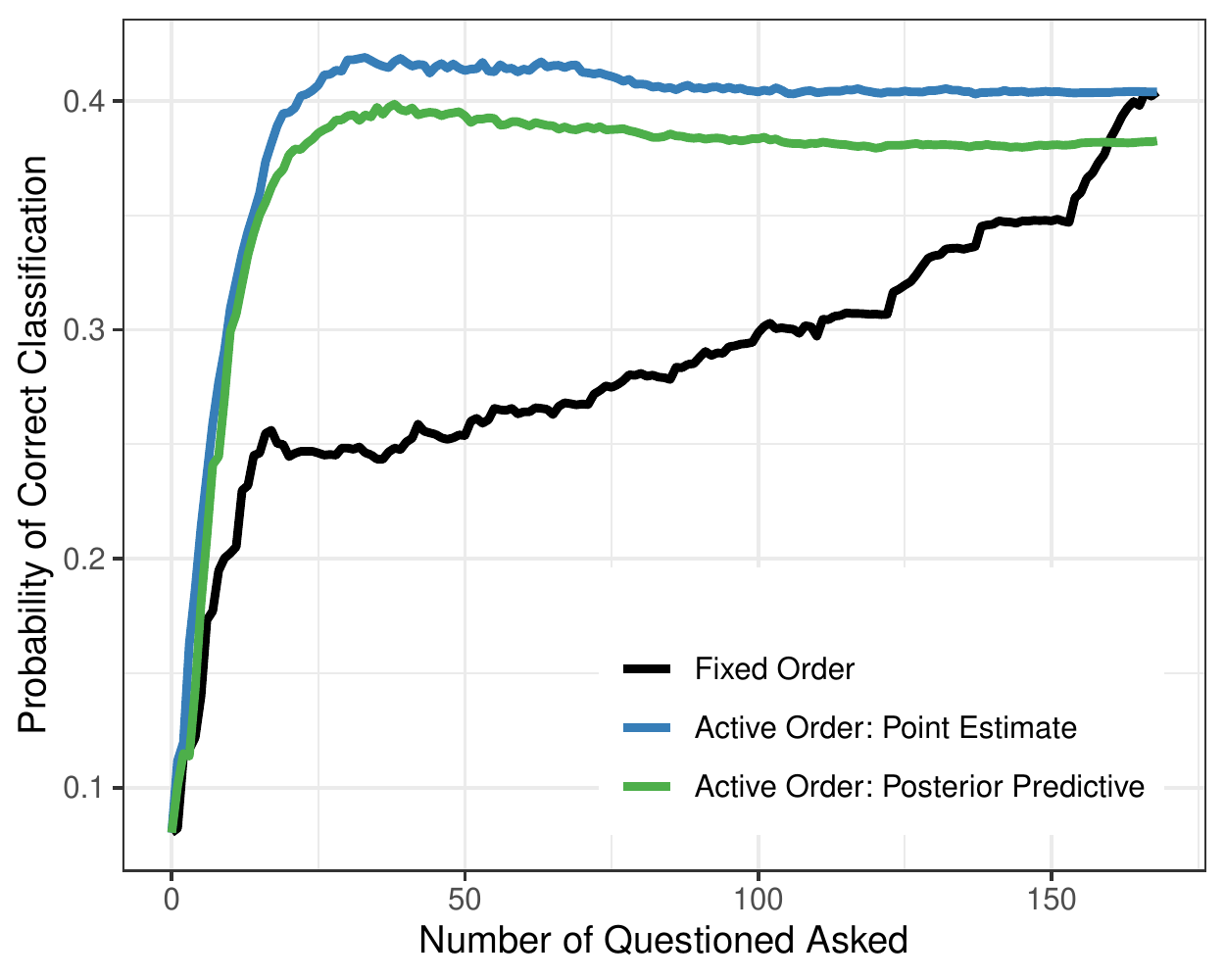}}
\end{figure}

Figure \ref{fig:phmrc1} shows the accuracy of the active questionnaire using point estimates and posterior predictive scores and compares them with the traditional questionnaire with fixed question order. Both strategies achieve a considerably higher classification accuracy compared to the static design after as few as $10$ questions. In fact, the overall accuracy of the active designs is optimal when around $30$ to $40$ questions have been asked, and then the accuracy slightly decreases as the questionnaire becomes longer. This is not surprising, as the analysis model in this experiment is likely overly simplistic and does not approximate the complex distribution of symptoms and causes well in the real data. However, while the design of better cause-of-death assignment algorithms remains an important research topic, the experiment clearly demonstrates that even with a simple analysis model, the active questionnaire design can lead to highly accurate cause-of-death classifications using only $1/4$ of the questions.

\begin{figure*}[htbp]
     \centering
\floatconts
  {fig:phmrc2}
  {\caption{Proportion of correctly classified deaths among deaths due to each cause using different stopping criteria. The causes are ordered by their sample size in the PHMRC data. The dotted horizontal lines correspond to accuracy from two widely adopted VA cause-of-death assignment algorithms using all symptoms. Red line: InSilicoVA \citep{mccormick2016}. Blue line: InterVA \citep{byass2019integrated}.}}
  {\includegraphics[width=\textwidth]{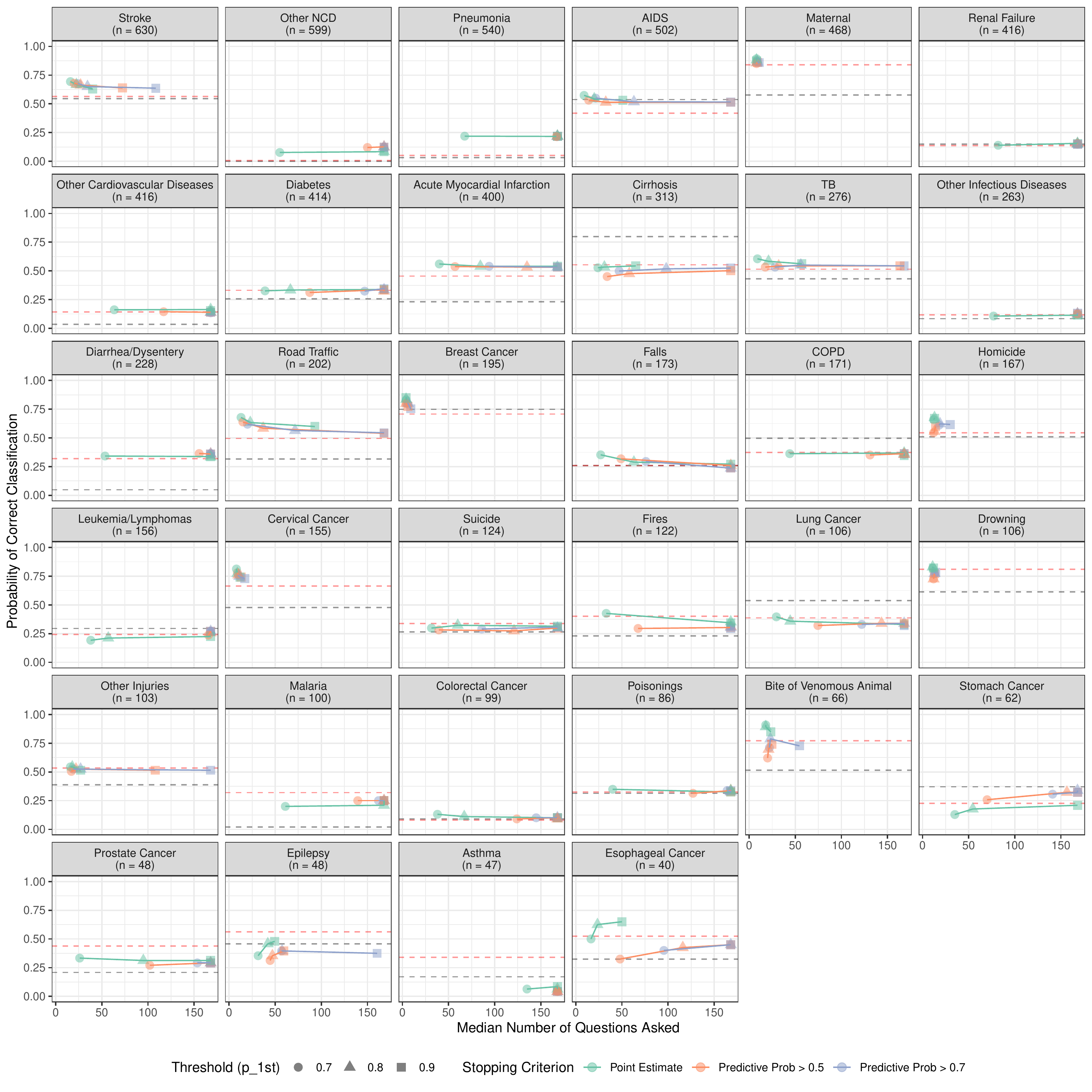}}
\end{figure*}

Lastly, we apply the adaptive early-stopping rule to the same cross-validation experiment. We consider $p_{1{\rm st}} = 0.7$, $0.8$, and $0.9$ and fix $d = 0.5$. We present additional sensitivity analysis using different values of $d$ in the supplementary materials. We again compare the stopping rules based on the point estimate and posterior predictive probabilities with $r = 0.5$ and $0.7$. Figure \ref{fig:phmrc2} examines the relationship between the proportion of correctly specified deaths when the questionnaire stops and the median number of questions asked, disaggregated by the true cause of death. For most causes, classification accuracy is not sensitive to the choice of thresholds we use, and only the length of the conducted questionnaire changes as the thresholds become more stringent. This is as expected from the observations in the first experiment that only a small number of questions is needed to achieve high classification accuracy. However, we note that for causes such as maternal deaths, breast cancer, cervical cancer, drowning, homicide, and bite of venomous animals, the active questionnaire design achieves high classification accuracy quickly regardless of stopping criterion. This is likely due to the strong associations between these causes and a small number of key symptoms. This observation is highly useful in practice as it allows interviewers to terminate the VA interview considerably sooner when sufficient information has been collected to identify certain causes that are easier to classify. We also compare the performance  of the active questionnaire strategy with two widely adopted VA cause-of-death assignment methods, InSilicoVA \citep{mccormick2016} and InterVA \citep{byass2019integrated} in Figure \ref{fig:phmrc2}. The active questionnaire strategy is able to achieve comparable and usually higher accuracy compared to both of the state-of-the-art VA algorithms using the full dataset.

\section{Discussion}
In this paper, we introduced a novel active questionnaire design strategy for verbal autopsy surveys. We proposed a principled Bayesian formulation to estimate posterior predictive scores of questions based on the KL information of questions in the bank. Our approach takes into account the uncertainty from any probabilistic cause-of-death assignment models and can facilitate adaptive early stopping rules and incorporate the existing flow of the questionnaire. We demonstrated improved performance on cause-of-death classification with both fixed and adaptive lengths of the questionnaire. More broadly, while we focus on verbal autopsy surveys in this work, the same methodology can be applied to other medical and health surveys in resource-constrained settings. 

The proposed active questionnaire can be readily adopted in the field as electronic data collection through tablets is already the standard practice for VA field interviews \citep{world2022standard}. The computation of the analysis model only needs to be performed once or updated routinely in a separate backend before data collection. The only computation required for real-time question selection involves the computation of \equationref{eq;PWKLScore} given pre-computed parameter values, which can be efficiently implemented on existing tablets.
The choice of the analysis model is contextual and depends on the implemented pipeline of VA data analysis. We anticipate that field experiments are needed to determine the choice of tuning parameters for the early termination of the questionnaires, which is beyond the scope of this paper. 

While this work provides a novel approach to questionnaire design in VA studies, we acknowledge several limitations for the approach to be directly useful in practice. First, the analysis model we used in this paper is oversimplified compared to the more recent work in the literature \citep[e.g., ][]{tsuyoshikunihama2020, li2021bayesian, wu2021double}. As a result, while we observe an advantage in the posterior predictive scores in determining early stopping rules, the overall classification accuracy is not significantly different from active ordering based on only the point estimates of the parameters. Combining the active design with more complex analysis models is beyond the scope of this work and is an important future direction. Second, there is extensive domain knowledge on the relationship and logic behind the questions on VA surveys. Such information may allow researchers to construct more useful penalty functions to regulate the flow of the active questionnaire and provide guidance on choosing the tuning parameter for the penalty.   

Several future directions of research could further address the methodological challenges of active questionnaire design for VAs. 
First, this work focuses on the situation where a cause-of-death assignment model has been chosen to analyze existing data. The active questionnaire strategy may be further improved to account for more than one analysis model. Ensemble prediction has been shown to improve the performance of cause-of-death classification \citep{datta2021} and could lead to active questionnaire designs and stopping rules that are more robust to model misspecification. 
Second, the recent work on domain adaptive VA algorithms illustrates that VA data are typically heterogeneous across different populations and the model parameters fitted with data from one population may not lead to good predictions for another population \citep{li2021bayesian, wu2021double}. The proposed adaptive design could be adapted to also account for this additional layer of uncertainty. 
Third, while we focus on the task of optimizing questionnaire design for future data collection given models estimated on existing data, it remains an open question how to efficiently combine and jointly analyze data collected via different adaptive and traditional instruments. 
Finally, it is an important practical research area to understand how VA interviewees respond to questionnaires with different orders and the potential impact of instruments on data quality. Furthermore, VA questions have different emotional burdens on the respondents, take varying amounts of time to conduct, and are subject to different types of bias across populations. It is also important to quantify the differential cost of each question to better understand the trade-off between classification accuracy and the cost of the interview. We leave these directions for future research.

\acks{The authors thank the anonymous reviewers for thoughtful feedback and discussion. The authors are especially grateful to Rrita Zejnullahi for her helpful comments during the preparation of this work. 
Research reported in this publication was supported by the Hellman Fellowship, Bill and Melinda Gates Foundation, the Eunice Kennedy Shriver National Institute of Child Health and Human Development (NICHD) under award number R03HD110962, and the National Institute of Mental Health of the National Institutes of Health under award number DP2MH122405. Partial support for this research came from an NICHD research infrastructure grant, P2C HD042828, to the Center for Studies in Demography \& Ecology at the University of Washington.}

% References
\bibliography{bib/VerbalAutopsy}

\clearpage
\onecolumn
\appendix

\section{Additional results for the synthetic data examples}

Tables \ref{tab;table_stop_correct_mean_full} and \ref{tab;table_stop_misspecified_mean_full} contain more extensive results on the classification accuracy and questionnaire length for the synthetic data under the correctly specified model and misspecified models, respectively. 

\begin{table}[!h]
\caption{Classification accuracy and questionnaire length for the synthetic data under the correctly specified model. Median and $5$th and $95$th percentiles of the questionnaire length are shown for each stopping rule conditions.\label{tab;table_stop_correct_mean_full}} 
\begin{center}
\begin{tabular}{cclcccc}
\hline\hline
\multicolumn{1}{c}{$p_{\rm 1st}$}&\multicolumn{1}{c}{$d$}&\multicolumn{1}{c}{Stopping Rule}&\multicolumn{1}{c}{Acc}&\multicolumn{1}{c}{Median}&\multicolumn{1}{c}{Lower}&\multicolumn{1}{c}{Upper}\tabularnewline
\hline
0.8&0.75&Point Est&0.84& 5& 3&14\tabularnewline
&&Pred $r=0.5$&0.92&10& 4&50\tabularnewline
&&Pred $r=0.7$&0.95&13& 6&50\tabularnewline
\hline 0.8&0.5&Point Est&0.85& 5& 3&14\tabularnewline
&&Pred $r=0.5$&0.92&10& 4&50\tabularnewline
&&Pred $r=0.7$&0.95&14& 6&50\tabularnewline
\hline 0.8&0.25&Point Est&0.88& 6& 3&16\tabularnewline
&&Pred $r=0.5$&0.94&10& 4&50\tabularnewline
&&Pred $r=0.7$&0.96&15& 6&50\tabularnewline
\hline 0.8&0&Point Est&0.96& 8& 5&35\tabularnewline
&&Pred $r=0.5$&0.96&13& 6&50\tabularnewline
&&Pred $r=0.7$&0.96&20& 7&50\tabularnewline
\hline 0.9&0.75&Point Est&0.95& 7& 5&17\tabularnewline
&&Pred $r=0.5$&0.95&12& 6&50\tabularnewline
&&Pred $r=0.7$&0.96&17& 7&50\tabularnewline
\hline 0.9&0.5&Point Est&0.95& 7& 5&18\tabularnewline
&&Pred $r=0.5$&0.95&12& 6&50\tabularnewline
&&Pred $r=0.7$&0.96&18& 7&50\tabularnewline
\hline 0.9&0.25&Point Est&0.95& 7& 5&24\tabularnewline
&&Pred $r=0.5$&0.96&12& 6&50\tabularnewline
&&Pred $r=0.7$&0.96&18& 7&50\tabularnewline
\hline 0.9&0&Point Est&0.97&10& 7&50\tabularnewline
&&Pred $r=0.5$&0.96&16& 8&50\tabularnewline
&&Pred $r=0.7$&0.96&23& 8&50\tabularnewline
\hline 0.95&0.75&Point Est&0.97& 9& 6&24\tabularnewline
&&Pred $r=0.5$&0.96&13& 7&50\tabularnewline
&&Pred $r=0.7$&0.96&18& 8&50\tabularnewline
\hline 0.95&0.5&Point Est&0.97& 9& 6&35\tabularnewline
&&Pred $r=0.5$&0.96&13& 7&50\tabularnewline
&&Pred $r=0.7$&0.96&20& 8&50\tabularnewline
\hline 0.95&0.25&Point Est&0.97&10& 6&50\tabularnewline
&&Pred $r=0.5$&0.96&15& 7&50\tabularnewline
&&Pred $r=0.7$&0.97&22& 8&50\tabularnewline
\hline 0.95&0&Point Est&0.98&12& 8&50\tabularnewline
&&Pred $r=0.5$&0.96&20& 8&50\tabularnewline
&&Pred $r=0.7$&0.97&26& 9&50\tabularnewline
\hline
\end{tabular}\end{center}
\end{table}

\begin{table}[!h]
\caption{Classification accuracy and questionnaire length for the synthetic data under the misspecified model. Median and $5$th and $95$th percentiles of the questionnaire length are shown for each stopping rule conditions.\label{tab;table_stop_misspecified_mean_full}} 
\begin{center}
\begin{tabular}{cclcccc}
\hline\hline
\multicolumn{1}{c}{$p_{\rm 1st}$}&\multicolumn{1}{c}{$d$}&\multicolumn{1}{l}{Stopping Rule}&\multicolumn{1}{c}{Acc}&\multicolumn{1}{c}{Median}&\multicolumn{1}{c}{Lower}&\multicolumn{1}{c}{Upper}\tabularnewline
\hline
0.8&0.75&Point Est&0.86& 4& 3&11\tabularnewline
&&Pred $r=0.5$&0.88& 4& 3&18\tabularnewline
&&Pred $r=0.7$&0.94& 5& 3&25\tabularnewline
\hline 0.8&0.5&Point Est&0.88& 5& 3&11\tabularnewline
&&Pred $r=0.5$&0.88& 4& 3&18\tabularnewline
&&Pred $r=0.7$&0.94& 5& 3&27\tabularnewline
\hline 0.8&0.25&Point Est&0.90& 5& 3&11\tabularnewline
&&Pred $r=0.5$&0.90& 4& 3&22\tabularnewline
&&Pred $r=0.7$&0.94& 6& 3&30\tabularnewline
\hline 0.8&0&Point Est&0.96& 6& 5&17\tabularnewline
&&Pred $r=0.5$&0.94& 6& 4&30\tabularnewline
&&Pred $r=0.7$&0.99& 8& 5&50\tabularnewline
\hline 0.9&0.75&Point Est&0.94& 6& 4&12\tabularnewline
&&Pred $r=0.5$&0.96& 6& 4&23\tabularnewline
&&Pred $r=0.7$&0.98& 7& 4&50\tabularnewline
\hline 0.9&0.5&Point Est&0.94& 6& 4&13\tabularnewline
&&Pred $r=0.5$&0.96& 6& 4&23\tabularnewline
&&Pred $r=0.7$&0.98& 7& 4&50\tabularnewline
\hline 0.9&0.25&Point Est&0.94& 6& 4&16\tabularnewline
&&Pred $r=0.5$&0.96& 6& 4&23\tabularnewline
&&Pred $r=0.7$&0.99& 7& 4&50\tabularnewline
\hline 0.9&0&Point Est&0.96& 7& 6&20\tabularnewline
&&Pred $r=0.5$&0.98& 8& 5&42\tabularnewline
&&Pred $r=0.7$&0.99& 9& 5&50\tabularnewline
\hline 0.95&0.75&Point Est&0.96& 7& 5&16\tabularnewline
&&Pred $r=0.5$&0.98& 7& 5&33\tabularnewline
&&Pred $r=0.7$&0.98& 8& 5&50\tabularnewline
\hline 0.95&0.5&Point Est&0.96& 7& 5&17\tabularnewline
&&Pred $r=0.5$&0.98& 7& 5&33\tabularnewline
&&Pred $r=0.7$&0.99& 9& 5&50\tabularnewline
\hline 0.95&0.25&Point Est&0.96& 7& 5&17\tabularnewline
&&Pred $r=0.5$&0.98& 8& 5&34\tabularnewline
&&Pred $r=0.7$&0.99& 9& 5&50\tabularnewline
\hline 0.95&0&Point Est&0.98& 8& 6&25\tabularnewline
&&Pred $r=0.5$&0.99& 8& 6&50\tabularnewline
&&Pred $r=0.7$&0.99&10& 6&50\tabularnewline
\hline
\end{tabular}\end{center}
\end{table}

\section{Additional results for the PHMRC data example}

\subsection{Descriptive statistics of the PHMRC dataset}

Table \ref{tab:sample-size} summarizes the sample size of each cause of death in the PHMRC data.

\begin{table}[htbp]
\centering
\begin{tabular}{rr}
  \hline
 Cause of death& Sample Size \\
  \hline
Stroke & 630 \\
  Other Non-communicable Diseases (NCD) & 599 \\
  Pneumonia & 540 \\
  AIDS & 502 \\
  Maternal & 468 \\
  Other Cardiovascular Diseases & 416 \\
  Renal Failure & 416 \\
  Diabetes & 414 \\
  Acute Myocardial Infarction & 400 \\
  Cirrhosis & 313 \\
  TB & 276 \\
  Other Infectious Diseases & 263 \\
  Diarrhea/Dysentery & 228 \\
  Road Traffic & 202 \\
  Breast Cancer & 195 \\
  Falls & 173 \\
  COPD & 171 \\
  Homicide & 167 \\
  Leukemia/Lymphomas & 156 \\
  Cervical Cancer & 155 \\
  Suicide & 124 \\
  Fires & 122 \\
  Drowning & 106 \\
  Lung Cancer & 106 \\
  Other Injuries & 103 \\
  Malaria & 100 \\
  Colorectal Cancer &  99 \\
  Poisonings &  86 \\
  Bite of Venomous Animal &  66 \\
  Stomach Cancer &  62 \\
  Epilepsy &  48 \\
  Prostate Cancer &  48 \\
  Asthma &  47 \\
  Esophageal Cancer &  40 \\
   \hline
\end{tabular}
\caption{Sample size of each cause of death in the PHMRC dataset.}
\label{tab:sample-size}
\end{table}

\subsection{More results of different stopping rules}
Figures \ref{fig:phmrc-sup-1} to \ref{fig:phmrc-sup-5} show more detailed results for the proportion of correctly classified deaths among deaths due to each cause of death using different stopping criteria.

\begin{figure*}[htbp]
     \centering
\floatconts
  {fig:phmrc-sup-1}
  {\caption{Proportion of correctly classified deaths among deaths due to each cause using different stopping criteria. The causes are ordered by their sample size in the PHMRC data.}}
  {\includegraphics[width=\textwidth]{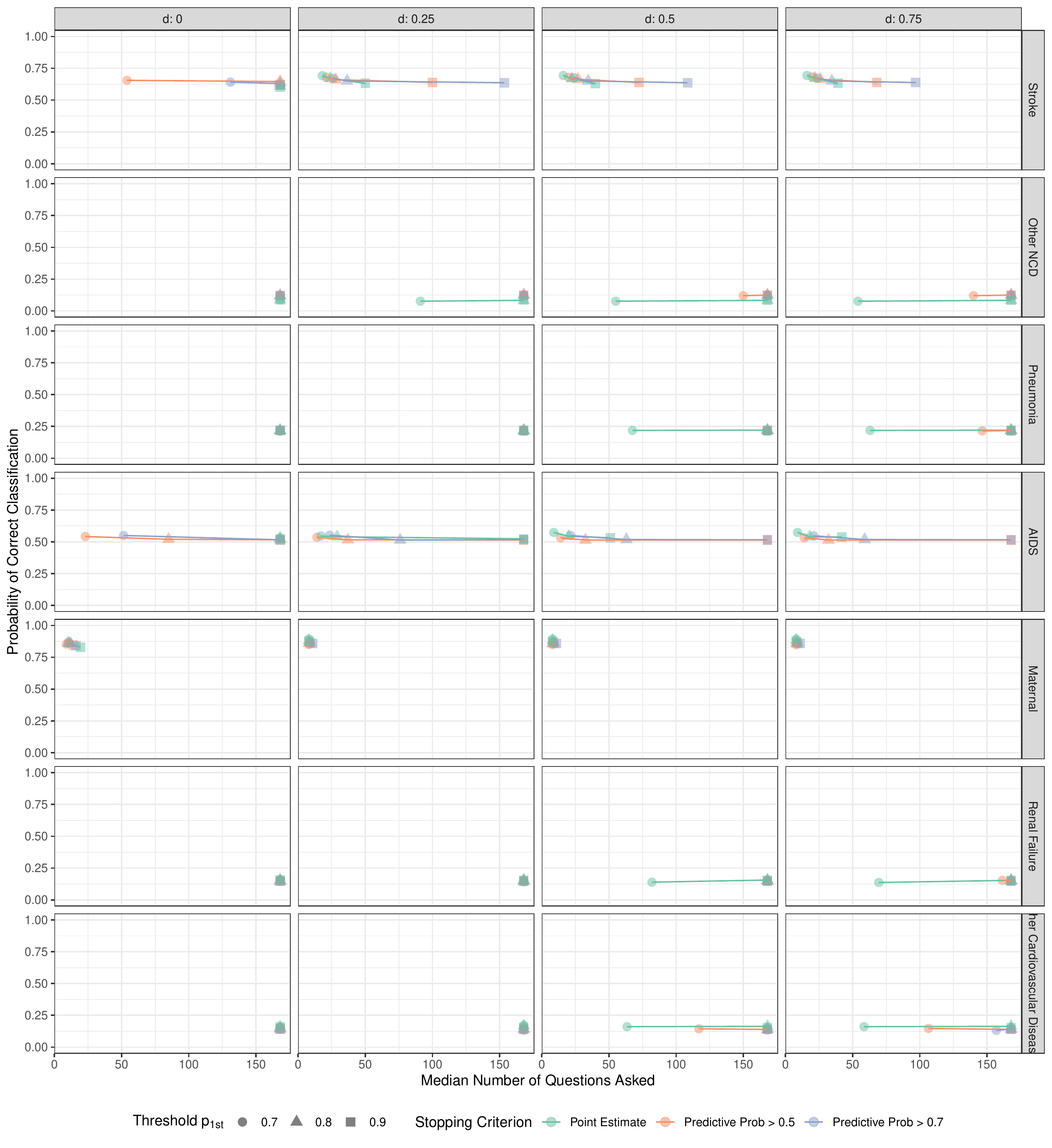}}
\end{figure*}

\begin{figure*}[htbp]
     \centering
\floatconts
  {fig:phmrc-sup-2}
  {\caption{(Continued) Proportion of correctly classified deaths among deaths due to each cause using different stopping criteria. The causes are ordered by their sample size in the PHMRC data.}}
  {\includegraphics[width=\textwidth]{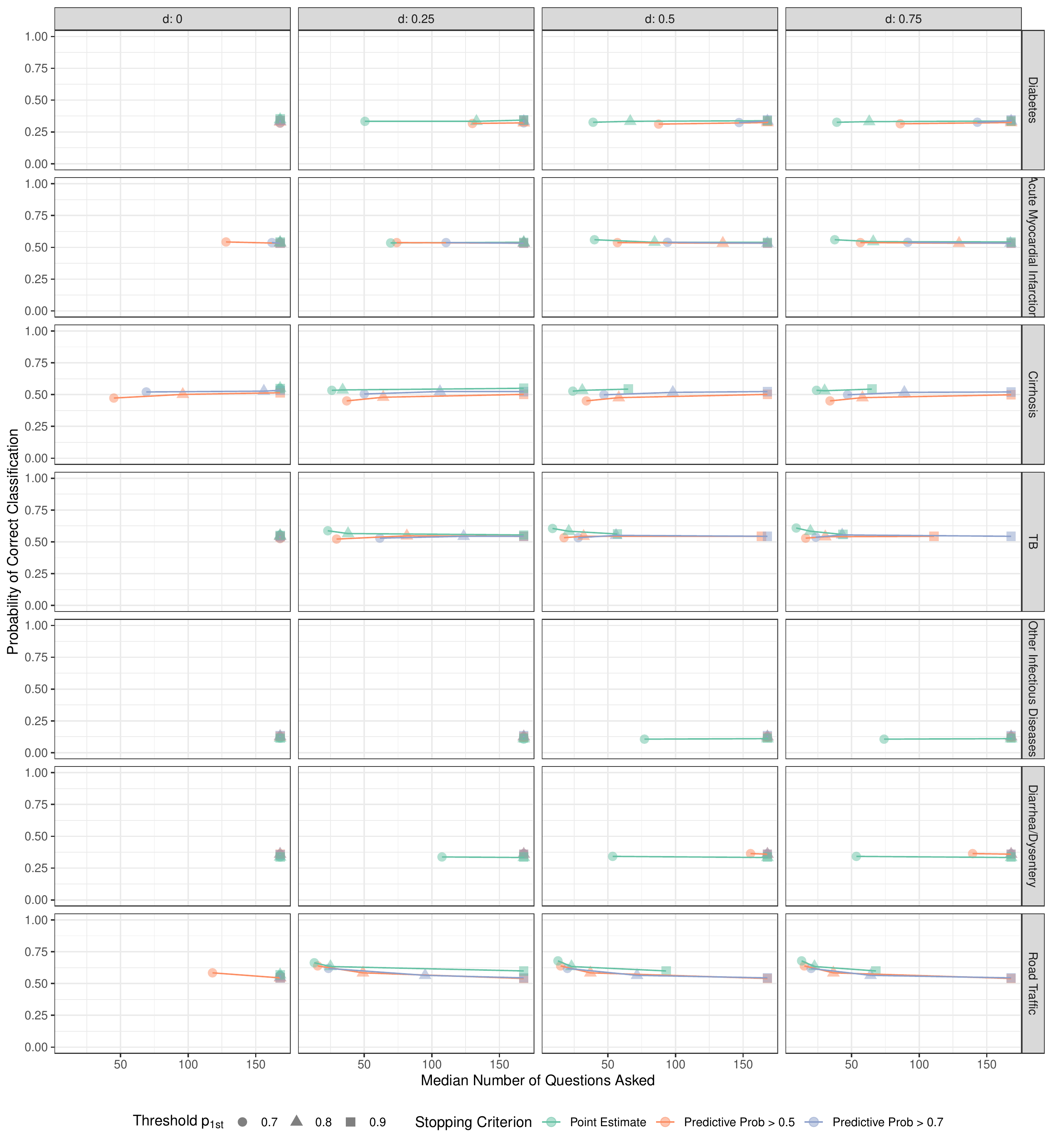}}
\end{figure*}

\begin{figure*}[htbp]
     \centering
\floatconts
  {fig:phmrc-sup-3}
  {\caption{(Continued) Proportion of correctly classified deaths among deaths due to each cause using different stopping criteria. The causes are ordered by their sample size in the PHMRC data.}}
  {\includegraphics[width=\textwidth]{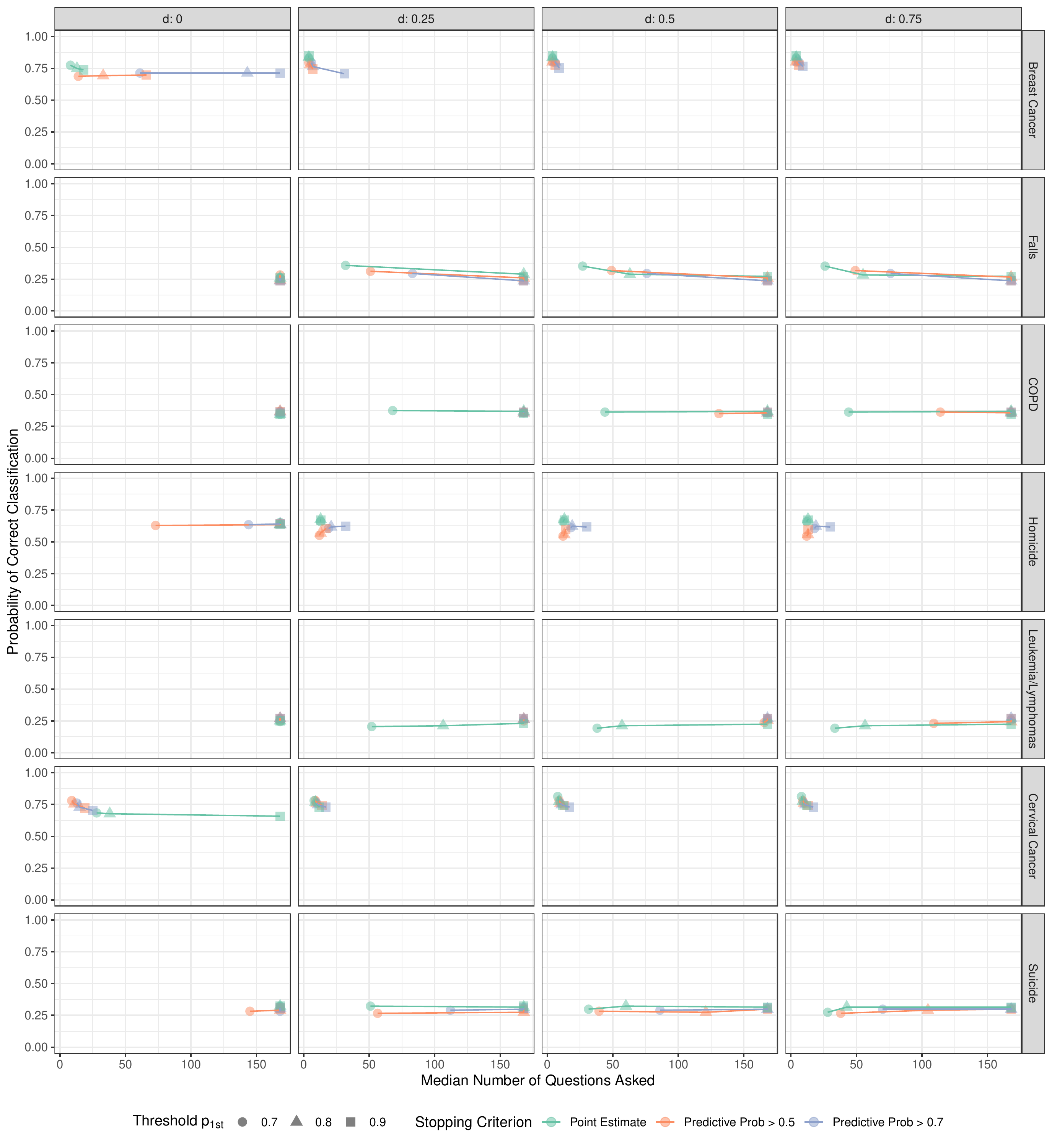}}
\end{figure*}

\begin{figure*}[htbp]
     \centering
\floatconts
  {fig:phmrc-sup-4}
  {\caption{(Continued) Proportion of correctly classified deaths among deaths due to each cause using different stopping criteria. The causes are ordered by their sample size in the PHMRC data.}}
  {\includegraphics[width=\textwidth]{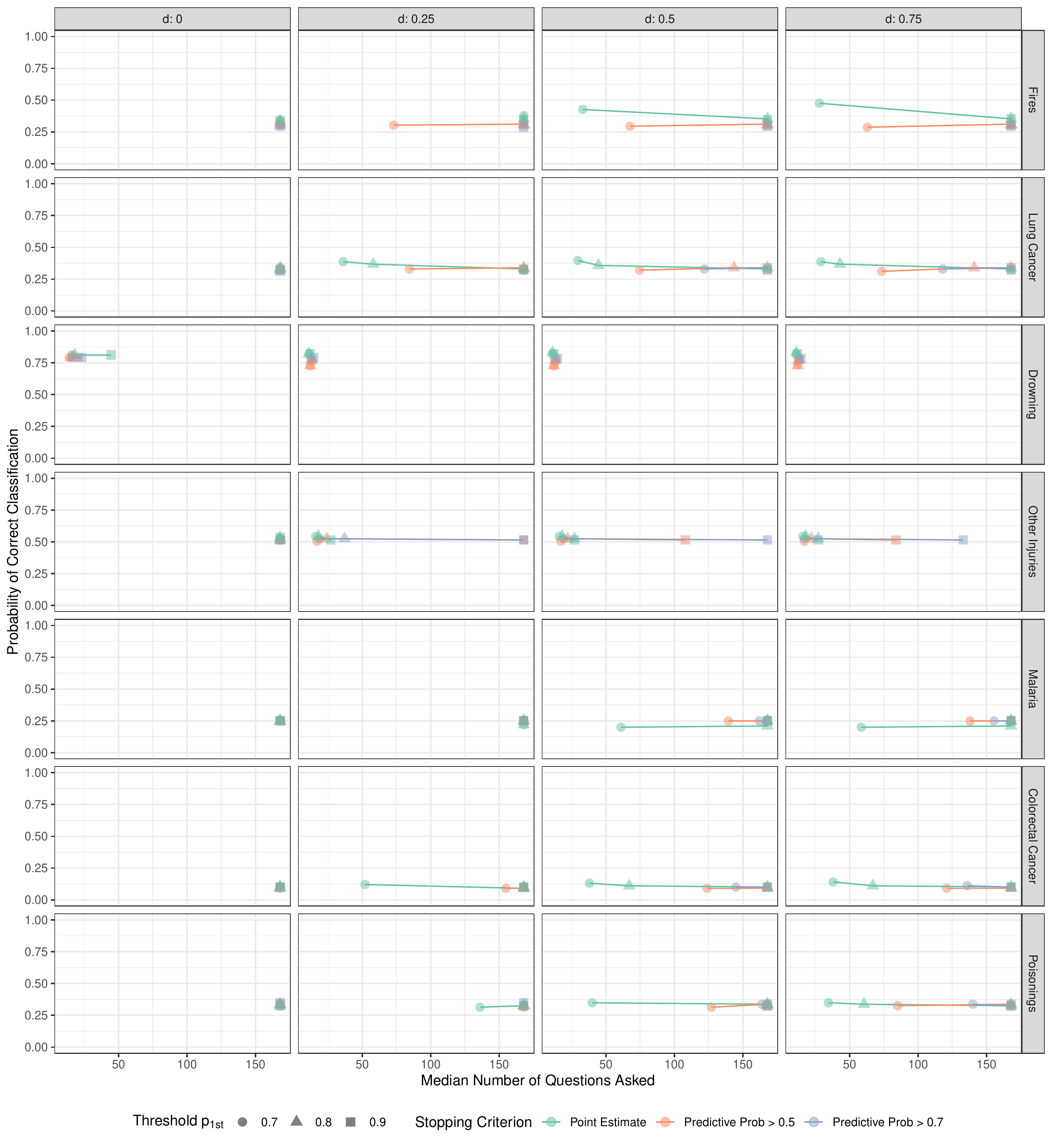}}
\end{figure*}

\begin{figure*}[htbp]
     \centering
\floatconts
  {fig:phmrc-sup-5}
  {\caption{(Continued) Proportion of correctly classified deaths among deaths due to each cause using different stopping criteria. The causes are ordered by their sample size in the PHMRC data.}}
  {\includegraphics[width=\textwidth]{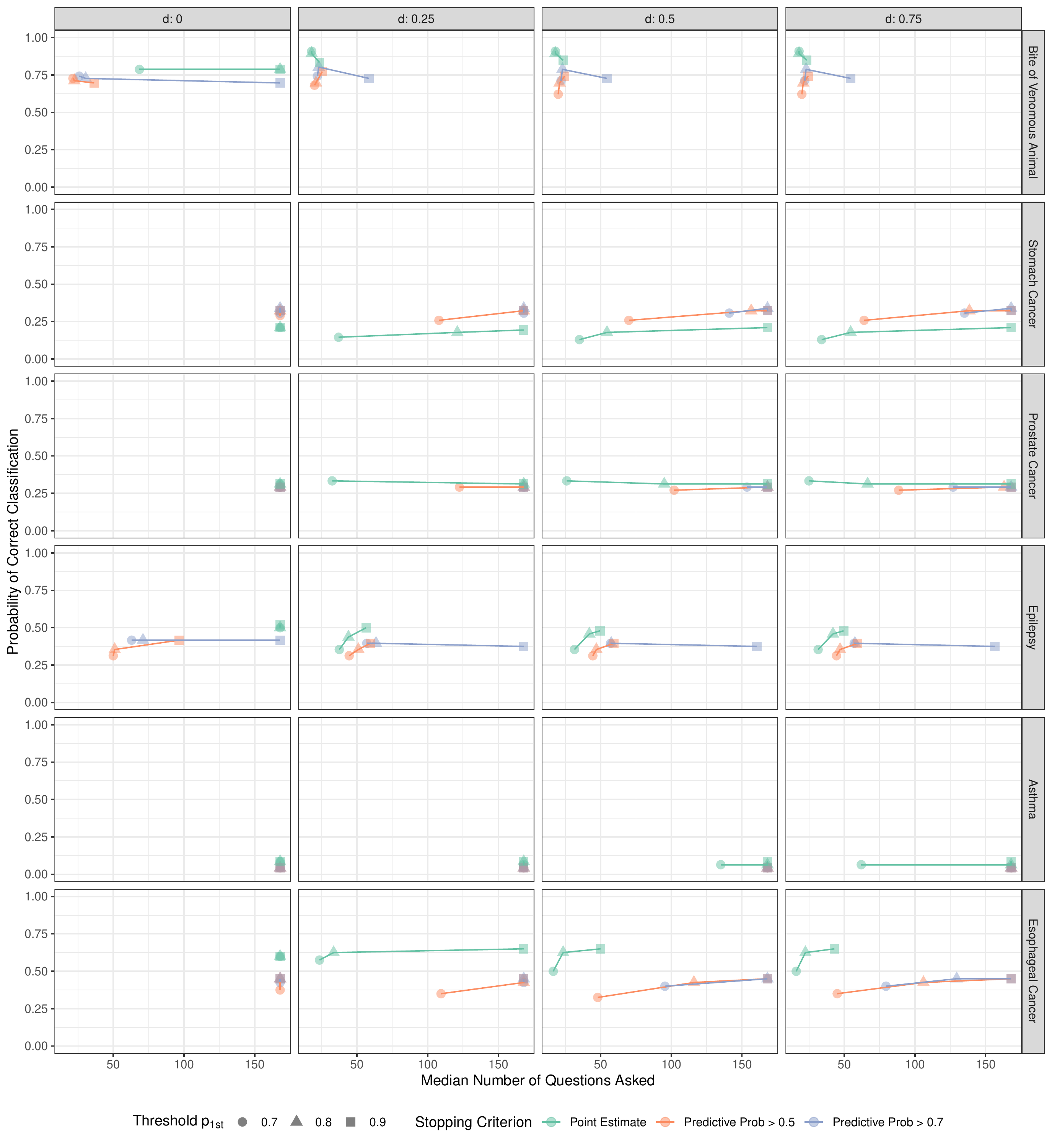}}
\end{figure*}
\end{document}